\documentclass[aps,twocolumn,superscriptaddress, groupedaddress,reprint]{revtex4}
\usepackage{graphicx}
\usepackage{setspace}
\usepackage{color}
\usepackage{float}
\usepackage{siunitx}
\usepackage{bm}
\usepackage{amsmath,amssymb}
\newcommand{\bea}{\begin{eqnarray}}
\newcommand{\eea}{\end{eqnarray}}
\newcommand{\bes}{\begin{subequations}}
	\newcommand{\ees}{\end{subequations}}

\usepackage{stackengine}

\begin{document}
	\title{Reviving Modulational Instability with  Third-Order Dispersion}
	\author{K. Tamilselvan}
	\affiliation{Nonlinear Waves Research Lab, PG and Research Department of Physics, Bishop Heber College, Tiruchirappalli-620 017, India}
	\author{A. Govindarajan}
\affiliation{Centre for Nonlinear Dynamics, School of Physics, Bharathidasan University, Tiruchirappalli-620 024, India}
	\author{T. Kanna}
	\affiliation{Nonlinear Waves Research Lab, PG and Research Department of Physics, Bishop Heber College, Tiruchirappalli-620 017, India}
	\author{M. Lakshmanan}
	\affiliation{Centre for Nonlinear Dynamics, School of Physics, Bharathidasan University, Tiruchirappalli-620 024, India}
	\author{P. Tchofo-Dinda}
	\affiliation{Laboratoire Interdisciplinaire Carnot de Bourgogne, UMR 6303 C.N.R.S., Universit\'e Bourgogne Franche-Comte\'e, 9 Av. A. Savary, BP 47870, F-21078 Dijon, France}
\begin{abstract}
	It is well-known that  third-order dispersion (TOD)
	never allows the continuous wave  of any nonlinear Schr\"odinger (NLS) type system to experience modulational instability (MI). Remarkably, we demonstrate a new kind of MI induced by TOD  with spatial dispersion accounting for Wannier exciton mass in a ZnCdSe/ZnSe semiconductor superlattice. We also numerically predict the existence of novel Akhmediev breathers and Peregrine solitons due to the nonlinear development of MI with TOD.
\end{abstract}

\maketitle
\section{Introduction}
In a broad range of physical systems, extremely weak perturbations can completely cause a phenomenon of chaotic energy localization, while in many other systems cooperative processes can develop at the microscopic level of matter and lead to an orderly structuring of the system.
However, the phenomenon of restructuring of the spectral density of energy in a system under the action of very weak perturbations can only occur in particular configurations  and specific operational conditions, which have aroused great interest for several decades.
Modulational instability (MI), also referred as Benjamin-Feir instability \cite{benjamin_feir_1967}, is undeniably one of  such well-known phenomena in which a continuous or quasi-continuous wave (CW) undergoes modulation of its amplitude or phase, because of its observation through fairly simple experimental setups in fields as varied as optics \cite{Tai86}, solid state physics \cite{Remoi99}, fluid mechanics \cite{benjamin_feir_1967}, plasma physics \cite{Akhtar2017}, and electrical lines \cite{Remoi99}.

 In optics, one of the simplest ways to generate MI is to propagate an intense laser light (called {\it pump power}) imposed with periodic perturbations
through a section of optical fiber in anomalous dispersion regime \cite{Tai86}.
Such a complex process of MI  is generally explained by
a phenomenon of four-wave mixing that can be written as
 $\omega_0+\omega_0 \rightarrow
(\omega_0+\Omega_\text{MI}) + (\omega_0-\Omega_\text{MI})$,
and interpreted as an inelastic collision during which two pump photons,  each having frequency $\omega_0$, are destroyed,
 while new Stokes and anti-Stokes photons are created, with frequencies $\omega_0-\Omega_\text{MI}$ and $\omega_0+\Omega_\text{MI}$, respectively.
Due to the {\it phase-matching condition}, the frequency detuning
of the sidebands $\Omega_\text{MI}$ depends not only on the pump power but also on some fiber parameters, which include the Kerr nonlinearity and even-order dispersions \cite{Potasek1987}. It is therefore well-known that odd-order dispersion coefficients, such as third-order dispersion (TOD), play no role in the MI process under the usual conditions of light propagation in optical fibers \cite{Potasek1987}. 
Such an interaction process can develop in a cascade and ultimately lead to complex wave dynamics like formation of Akhmediev breathers (ABs), Kuznetsov-Ma  breathers (KMBs),
and Peregrine solitons (PSs).  Subsequently, various nonlinear waves induced by MI  have been experimentally demonstrated in \cite{Baronio2017}, in addition to the generation of ultra-short pulses at high repetition rate (at tera-hertz frequency), which is also linked to  Fermi-Pasta-Ulam-Tsingou
 recurrence \cite{van}, and super-continuum generation \cite{Dudley}.

As it is well-known that optical fibers act as a great platform to
investigate multiple nonlinear phenomena, including MI in the framework of nonlinear optics \cite{Kivshar}. Similarly, semiconductor materials also share a suitable ground for exploring the generation of nonlinear waves in the vicinity of exciton-polaritons (EPs) \cite{Egorov, Biancalana}. The concept of EP can be defined by electronic excitations in semiconductors, which is then classified as Wannier-Mott exciton or Frenkel's exciton depending on their Coulomb interaction happening between electrons and holes. Unlike in optical fibers, here Kerr nonlinearity stems from the Coulomb interactions between optically excited states \cite{EPbook}. 
Such models can be governed by NLS-like equation by obeying slowly-varying envelope approximation (SVEA). The phenomenon of MI has also been intensively investigated in numerous types of EP materials which include microcavities \cite{Egorov,Biancalana,Smyrnov, EPbook2,EPbook3}. These models generally include a term of spatial dispersion (SD) which is proportional to the Wannier-exciton mass \cite{Biancalana}. Following the seminal work of \cite{Biancalana}, we propose  a higher-order model of dimensionless  NLS system which includes the role of TOD ($\delta_{3}$), which should be taken into account when the pulse operates near zero-dispersion regime \cite{Saka1}, with the Wannier exciton mass in a ZnCdSe/ZnSe semiconductor superlattice \cite{Biancalana, EPbook2,EPbook3, McDonald, McDonald1}: 
\bea\label{GNLH}
i\Psi_{z}+\Lambda~\Psi_{zz}+\frac{1}{2}~\Psi_{tt}-i \delta_{3}~\Psi_{ttt}+|\Psi|^2 \Psi=0,\quad
\eea
where  $\Psi$ is the complex envelope field, $z$ is the spatial coordinate  and  $t$ refers the retarded time. The TOD term is noted as $\delta_{3}=\beta_{3}/6 T_{0} |\beta_{2}|$, where $T_0$ is the typical pulse width and $\beta_j$ ($j=2,3$) indicate second and third order group velocity dispersions in physical units. Here the third term in \eqref{GNLH} is scaled to be one and assumed to operate in the anomalous dispersion regime. It has been shown that for $\Lambda>0$ the system (\ref{GNLH}) exhibits conventional bulk Wannier exciton resonances describing the dynamics of \textit{exciton-polariton} (EP) with positive effective mass in the ZnCdSe/ZnSe superlattice  \cite{Biancalana}. Note that there is also a special case when $\Lambda<0$ denoting a negative exciton mass which is not of interest here. In this paper, we consider the  positive effective mass in normalized unit of  $\Lambda$ which ranges from \emph{zero to unity} attributing to the cases of \emph{low effective mass (LEM)} to \emph{high effective mass (HEM)}, respectively. 
The last term in \eqref{GNLH} is the Kerr nonlinearity where the Kerr coefficient is scaled to be one. Yet so far, the study of MI remains unexplored in the context of NLS systems with TOD except for the work of Mussot \textit{et al.} \cite{Mussot} which dealt with the numerical global stability analysis  employing \emph{time-localized pulses} rather than the CW solution. To the best of our knowledge, for the first time, we show here the dynamics of MI for the higher-order NLS system \eqref{GNLH} with TOD describing the EP in ZnCdSe/ZnSe semiconductor superlattice.
\section{Modulational instability (MI) analysis}
\begin{figure}[t]\centering
	\includegraphics[width=0.5\linewidth]{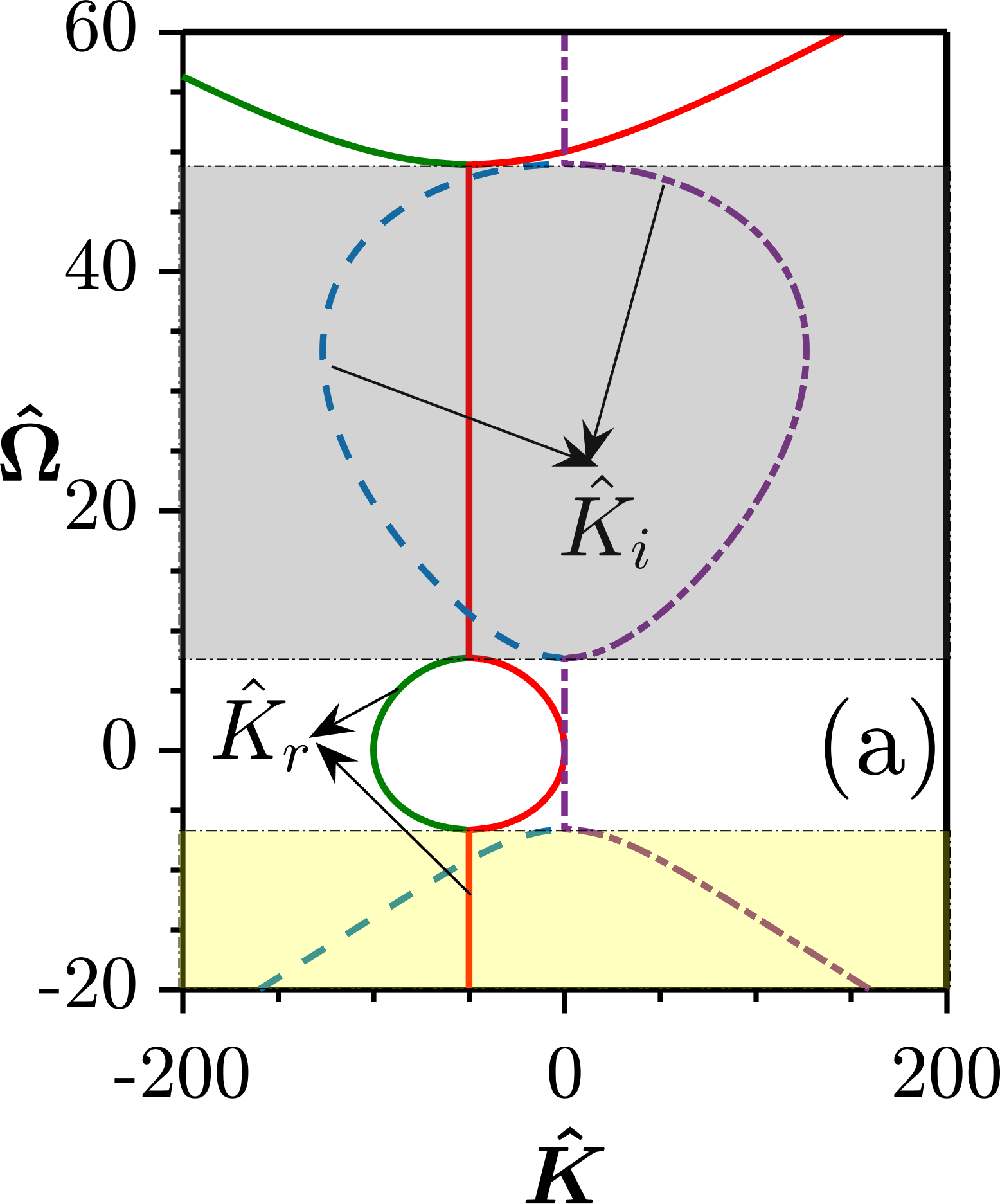}~
	\includegraphics[width=0.5\linewidth]{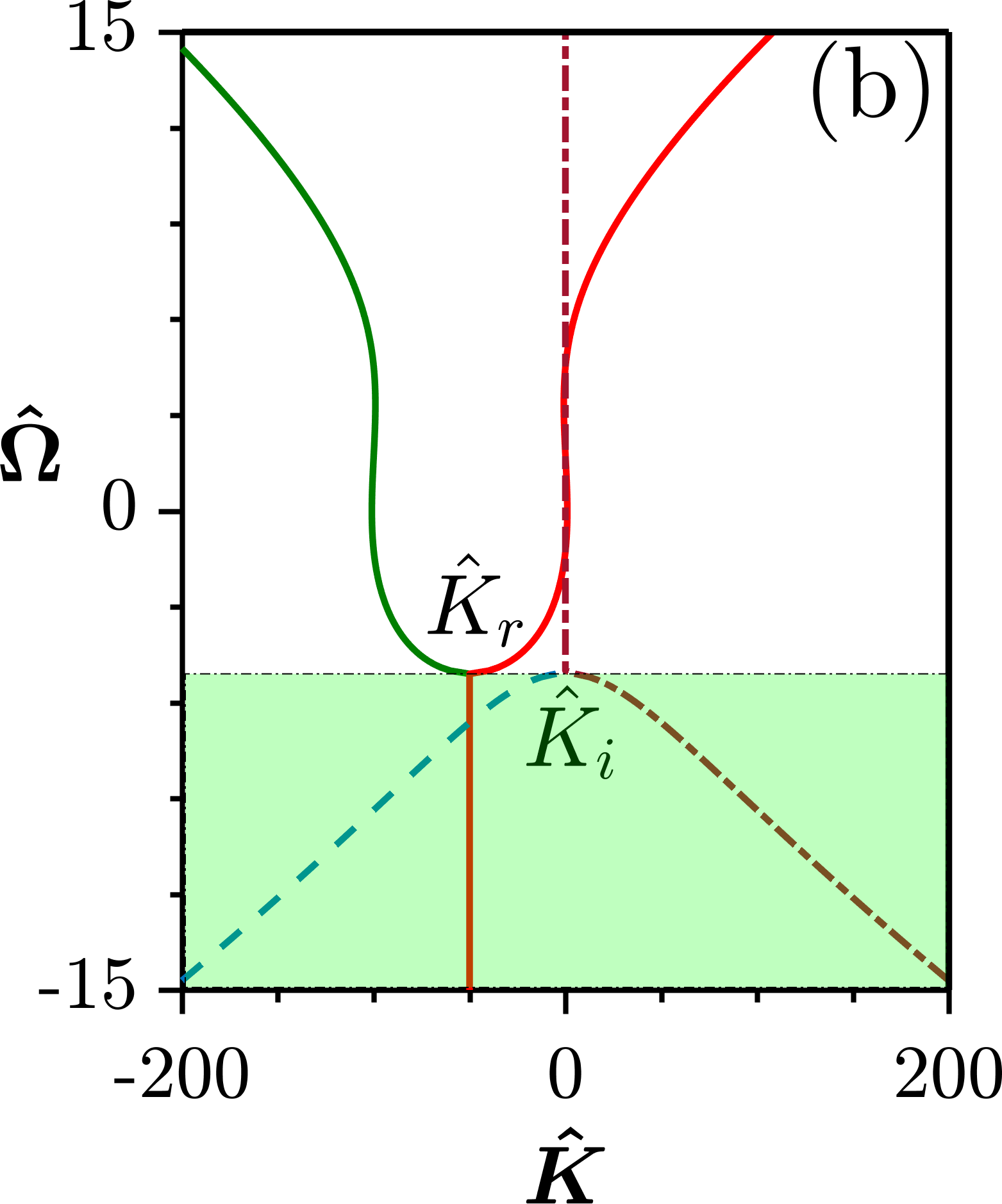}\\
	\caption{Dispersion relation for system with $\Lambda=0.01$. The parameters are (a) $\delta_{3}=0.01$ and (b) $\delta_{3}=0.1$ with $P=10$.}
	\label{fig:dispersion}
\end{figure}

Since the dispersion relation can reveal all possible interactions of light waves in the system considered, we first examine its characteristics. To this end,  we substitute the CW solution $q(z,t)=\sqrt{P} e^{i(\hat{K} z-\hat{\Omega} t)}$ into \eqref{GNLH}, where $P$ is the input peak power, $\hat{K}$ and $\hat{\Omega}$ are, respectively, the wave vector and frequency. This results in a nonlinear dispersion relation $\hat{K}(\hat{\Omega})=\frac{-1+\gamma\sqrt{1+4P\Lambda-2  \Lambda\hat{\Omega}^2+4\delta_{3}\Lambda\hat{\Omega}^3}}{2\Lambda}$, where $\gamma =\pm1$ defines, respectively, upper and lower branches of the dispersion curve. 
Figure \ref{fig:dispersion} shows quite unusual bands in contrast to the conventional ones, where one can observe a wide range of forbidden frequencies creating a photonic band gap which exists between two bands (one with the conventional dispersive band and other with a circular one) having real $\hat{K_r}$ values (gray shaded region) when $\delta_{3}$ is low (see Fig. \ref{fig:dispersion}(a)). In addition, at the edge and outside of the circular band, the wave vector becomes predominantly imaginary ($\hat{K_i}$) thereby not allowing any localized states (yellow shaded area). As $\delta_{3}$ is increased to 0.1 (Fig. \ref{fig:dispersion}(b)), the dispersion takes a new dimension by transforming into an elliptic curve with a real wave vector. Nevertheless, most of the negative frequencies ($\hat{\Omega}<-5$) becomes the forbidden one by not supporting any propagating wave. It is to noted that we have also studied the dispersion relation for the system with HEM ($\Lambda=1$) and observed that the range of forbidden frequencies is even more when compared to the system with LEM ($\Lambda=0.01$)  albeit they exhibit somewhat different dispersion curves (not shown here). These findings clearly suggest that the considered systems may support some novel kinds of localized states (as the dispersion relation of the present system is having analogy with the cryptographic system \cite{Enge,Fujioka}) which we leave for a detailed subsequent work.

\subsection{Linear stability analysis}
To explore the onset of MI of the \emph{steady-state solution} against an infinitesimal perturbation $a(z,t)$, we substitute (without loss of generality by setting $\hat{\Omega}=0$ in the CW solution)  $q(z,t)=(\sqrt{P}+a(z,t))\exp(i\hat{K}z)$ in \eqref{GNLH} and after linearizing, we obtain
\bea\label{linearizedEq}
\nu~a_{z}+\Lambda~a_{zz}+\frac{1}{2}~a_{tt}-i \delta_{3}~a_{ttt}+P (a+a^{*})=0,\qquad
\eea
where $\nu=\pm i \sqrt{4 \Lambda P+1}$. After expressing  the perturbed CW solution in terms of its Fourier components as $a(z,t)=a_{1}\exp(i(K z- \Omega t))+a_{2} \exp(-i(K z- \Omega t))$, where $a_{1,2}$ are arbitrary real constants, and $K$ and $\Omega$ are, respectively, complex wave number and frequency of the perturbation, we arrive at a set of homogeneous equations for $a_{1}$ and $a_{2}$. This set of characteristic equations admits four branches of solutions when $K(\Omega)$ satisfies the following relation
\bea\label{wavenumber}
K_{1-4}=\pm_{\epsilon_{1}}\frac{1}{2}\left({\sqrt{sgn(R_{1})|R_{1}|}}\pm_{\epsilon_{2}}\sqrt{sgn(R_{2})| R_{2}|}\right).
\eea
The expressions of $R_{1}$ and $R_{2}$ are given by $-2 \hat{\Upsilon}_{2}/3 \hat{\Upsilon}_{1}+\Gamma$ and $-4 \hat{\Upsilon}_{2}/3\hat{\Upsilon}_{1}-\Gamma \pm_{\epsilon_{3}}2\hat{\Upsilon}_{3}/$\\$(\hat{\Upsilon}_{1} \sqrt{2 \hat{\Upsilon}_{2}/(\hat{3 \Upsilon}_{1})+\Gamma})$, respectively. Here the parameters $\hat{\Upsilon}_{1,2,3,4}$, and $\Gamma$ are given by $\Lambda^{2}$,$-(1+6 P \Lambda - \Lambda \Omega^{2})$, $-2 \delta_{3} \sqrt{1+4 P \Lambda} \Omega^{3}$, $-P \Omega^{2}+\frac{\Omega^{4}}{4}-\delta_{3}^{2} \Omega^{6}$, and
$(2^{2/3}\upsilon_{1}\upsilon_{2}+(\upsilon_{1}+\sqrt{-4\upsilon_{2}^{3}+\upsilon_{1}^{2}})^{2/3})/3\hat{\Upsilon}_{1}[2(\upsilon_{1}+\sqrt{-4\upsilon_{2}^{3}+\upsilon_{1}^{2}})]^{1/3}$, respectively, in which $\upsilon_{1,2}$ can be expressed as $2 \hat{\Upsilon}_{2}^{3}+27 \hat{\Upsilon}_{1} \hat{\Upsilon}_{3}^{2}-72 \hat{\Upsilon}_{1} \hat{\Upsilon}_{2} \hat{\Upsilon}_{4}$ and
$\hat{\Upsilon}_{2}^{2}+12 \hat{\Upsilon}_{1}\hat{\Upsilon}_{4}$, accordingly. The four roots $K_{1}, K_{2}, K_{3}$ and $K_{4}$, respectively result for the distinct combinations $(+_{\epsilon_{1}},-_{\epsilon_{1}},-_{\epsilon_{3}})$, $(+_{\epsilon_{1}},+_{\epsilon_{1}},-_{\epsilon_{3}})$, $(-_{\epsilon_{1}},-_{\epsilon_{1}},+_{\epsilon_{3}})$, $(-_{\epsilon_{1}},+_{\epsilon_{1}},+_{\epsilon_{3}})$. From the dispersion relation shown in Eq. \eqref{wavenumber}, one can find the branches that become imaginary for the following choices (i) $sgn(R_{1})<0$ and  $sgn(R_{2})<0$, (ii) $sgn(R_{1})>0$ and  $sgn(R_{2})<0$, and (iii) $sgn(R_{1})<0$ and  $sgn(R_{2})>0$. Hence, the MI gain spectrum $G(\Omega)$ can be calculated as $G(\Omega)=2|Im(K_{max})|$, where $Im(K_{max})$ is the largest imaginary part of the eigenvalues. It should be stressed that the TOD influences the imaginary part of the eigenvalues which leads to the MI growth in the present system. Conversely, in the case of standard NLSE $(\Lambda=0)$, the TOD has no effect at all in the MI growth rate since the former appears in the real part of eigenvalues alone, which is given by, $K=\frac{1}{2}(2\delta_{3}\Omega^3\pm\sqrt{-4P\Omega^2+\Omega^4})$.

\begin{figure}[t]
	\centering
	\includegraphics[width=0.49\linewidth]{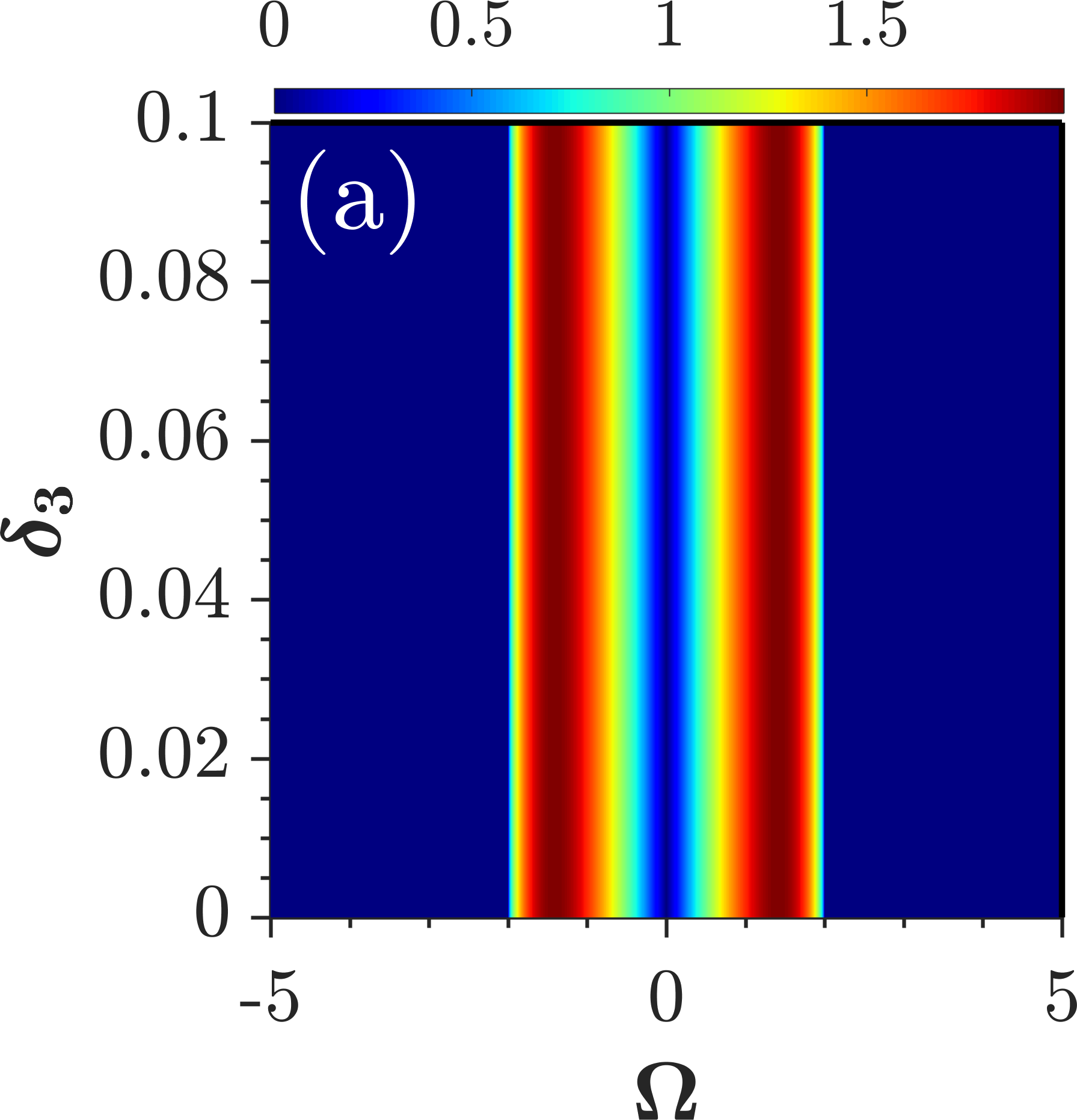}
	\includegraphics[width=0.49\linewidth]{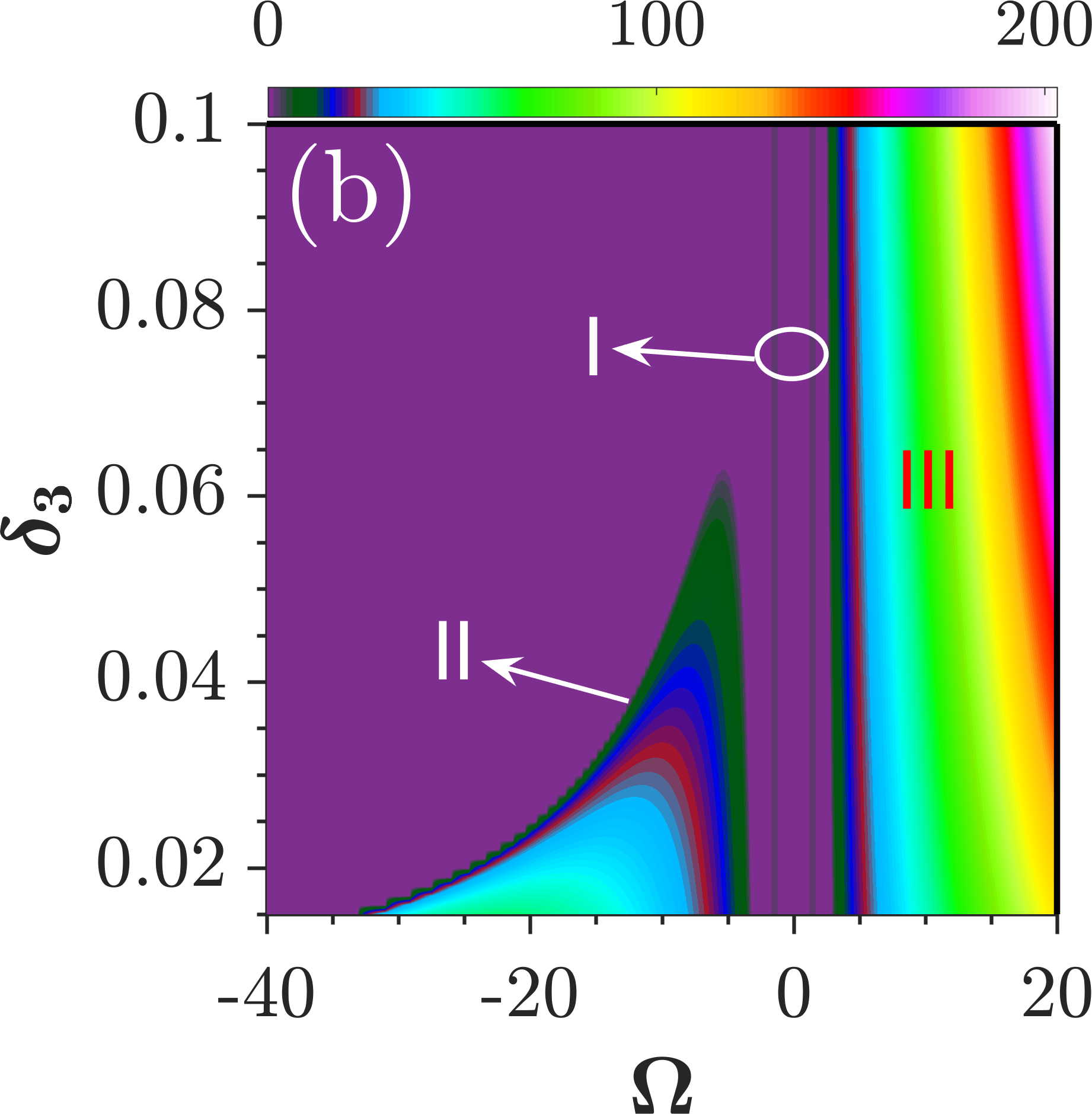}\\
	\includegraphics[width=0.49\linewidth]{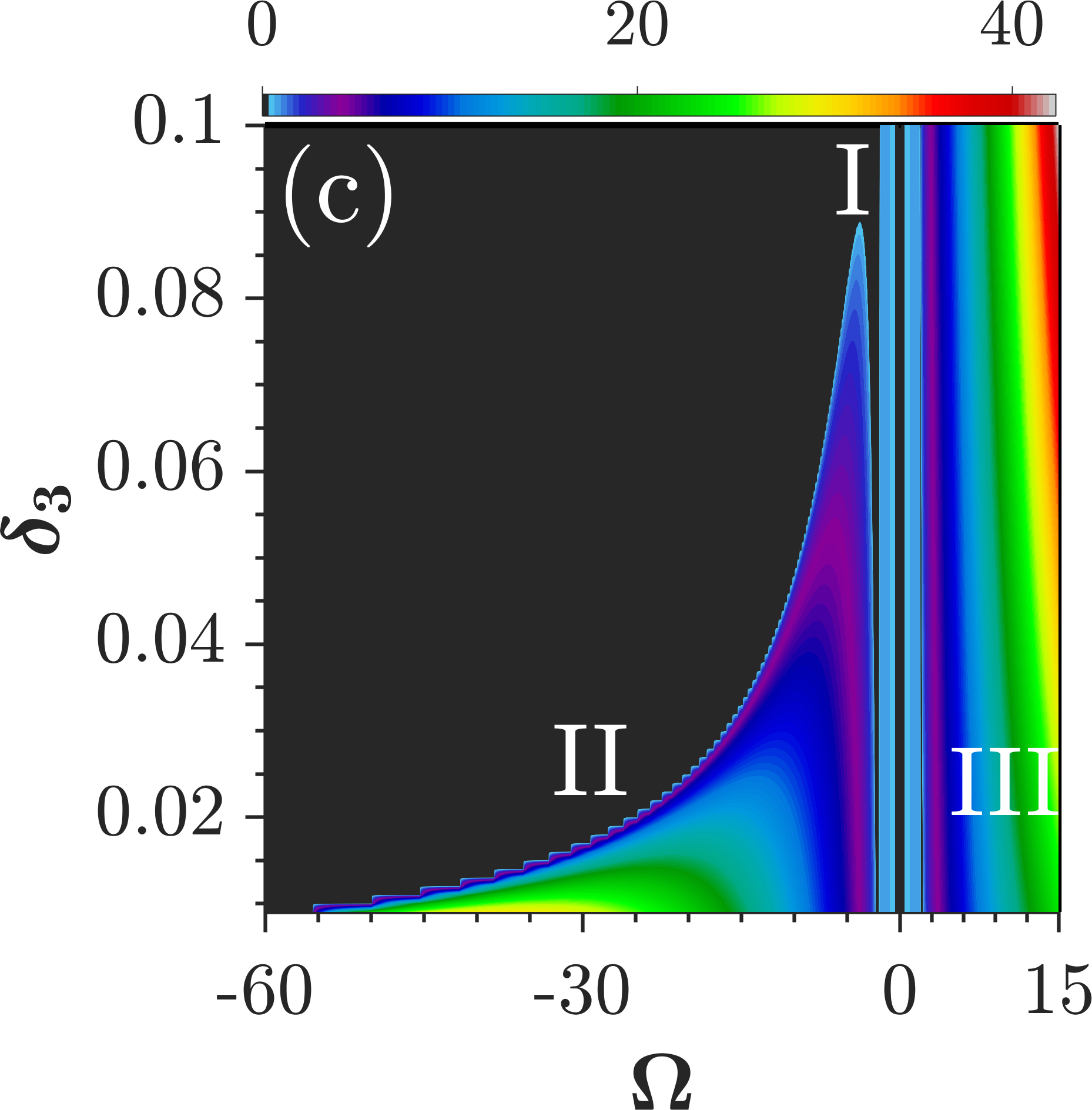}
	\includegraphics[width=0.49\linewidth]{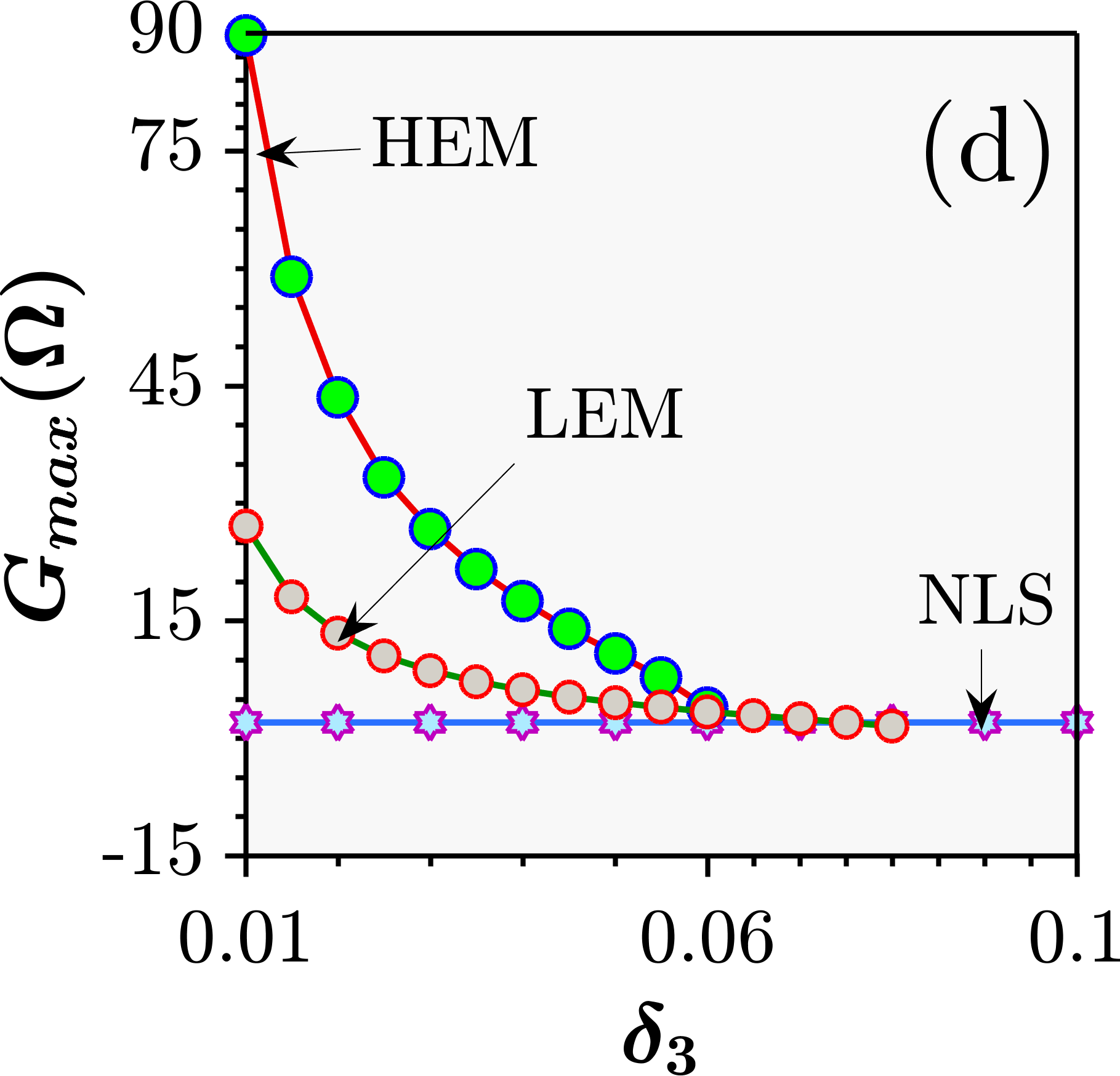}
	\caption{Role of TOD on the MI spectra for different (a) NLS ($\Lambda=0$), (b) system with $\Lambda=0.09$, and (c) system with $\Lambda=1$ systems. (d) Depicts the peak gain versus TOD for various systems. Other parameters are fixed as $P=10$.}
	\label{fig:TOD}
\end{figure}
The central outcome of this paper is demonstrated in Fig. \ref{fig:TOD}, where Fig.  \ref{fig:TOD}(a) reveals that the TOD remains completely oblivious to the MI gain spectra for the paraxial NLS system ($\Lambda=0$), which is further corroborated in Fig.~\ref{fig:TOD}(d) (refer to the blue solid line with hexagram markers). On the contrary, for the case of LEM ($\Lambda=0.09$) (see Fig. \ref{fig:TOD}(b)), we obtain a broad  MI spectrum owing to the role of TOD at Stokes frequency with a dramatic growth rate (see regime II). More precisely, the peak gain of the spectrum is increased by about a factor of 50 times compared to the conventional one (cf. Fig. \ref{fig:TOD}(b) and regime I). A further increase in $\delta_{3}$ makes the sidebands to experience a complete shifting towards the central perturbation frequency featuring a narrow spectrum with reduced peak gain. It is worthwhile to mention that such a manifestation of shifting is a typical signature of any frequency dependent parameter including the TOD effect \cite{Kodama}, though the latter exhibits an opposite MI dynamics compared to the walk-off. We hence affirm that the existence of seonecondary sidebands at the  Stokes frequency is purely due to the impact of TOD which is completely a new observation in the context of nonlinear optics. To substantiate this, we illustrate its maximum gain (traced at region II)  versus the TOD parameter as presented in Fig.~\ref{fig:TOD}(d) (green circles), where one can clearly observe that the peak gain of MI spectrum gets reduced  as the TOD is increased. In addition to this, at anti-Stokes frequency, there exists a continuous growth (instead of exponential growth) of unstable modes as an outcome of SD (region III), which has already been reported in our previous work \cite{Tamil2019}. Note that such an impact of SD exactly mimics the continuous growth resulting from the stimulated Raman scattering effect.  Further, we concentrate on the MI behavior in the case of HEM by setting the spatial dispersion parameter to be unity ($\Lambda=1$), see Fig. \ref{fig:TOD}(c). Even though the spectra obtained for this case are exactly same  as that of Fig. \ref{fig:TOD}(b) as expected, the major difference now is that the maximum gain is diminished about ten times  compared to the case of weak effective mass (cf. green and red circles in Fig.  \ref{fig:TOD}(d)). Also, the range of Stokes frequency at which the spectra exist induced by the TOD gets wide in addition to the increase in the value of TOD at which the system with HEM ($\Lambda=1$) experiences further unstable state ($\delta_{3}\approx0.089$).

\begin{figure}[t]
\centering
\includegraphics[width=0.49\linewidth]{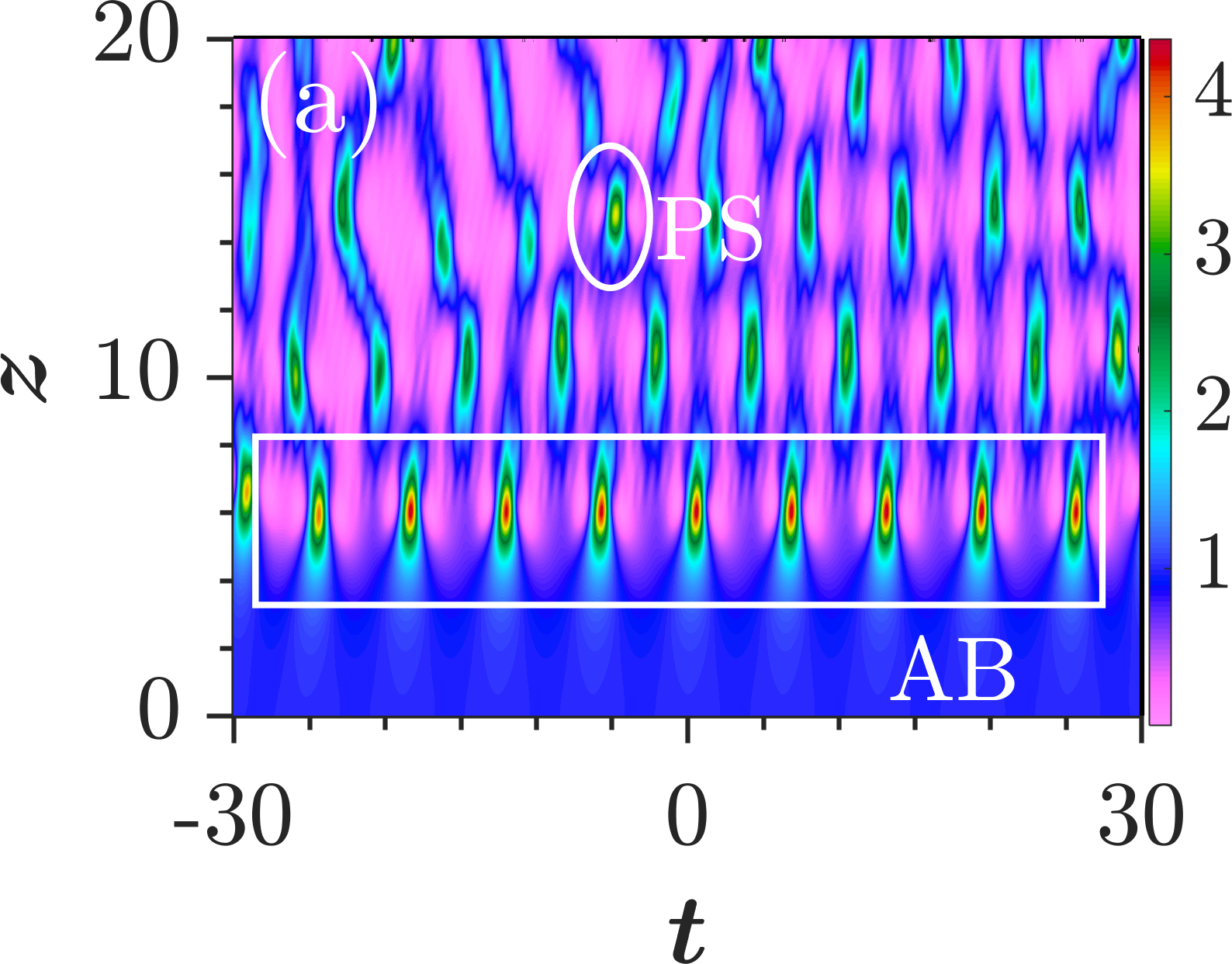}
	\includegraphics[width=0.48\linewidth]{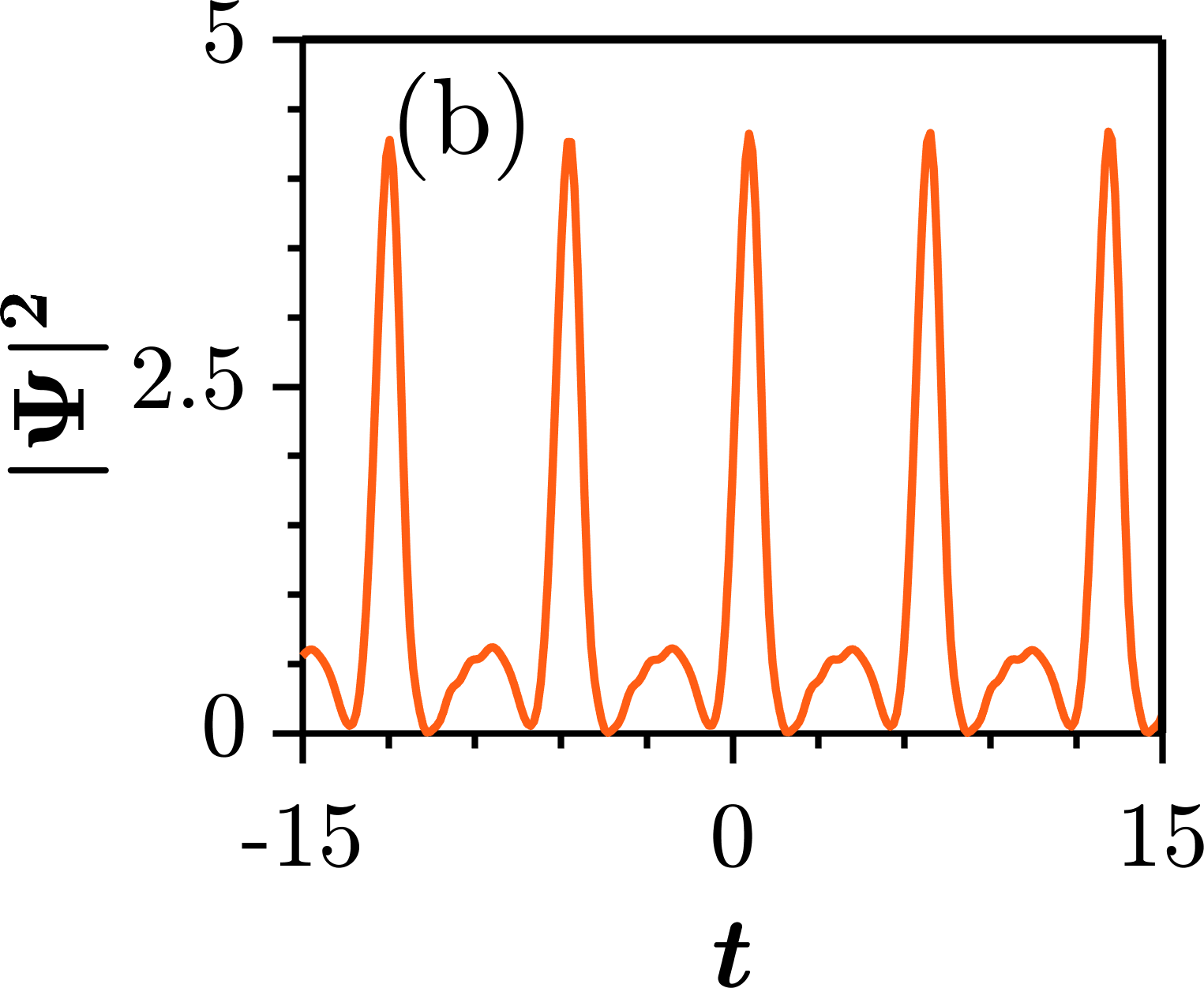}\\\centering
	\includegraphics[width=0.49\linewidth]{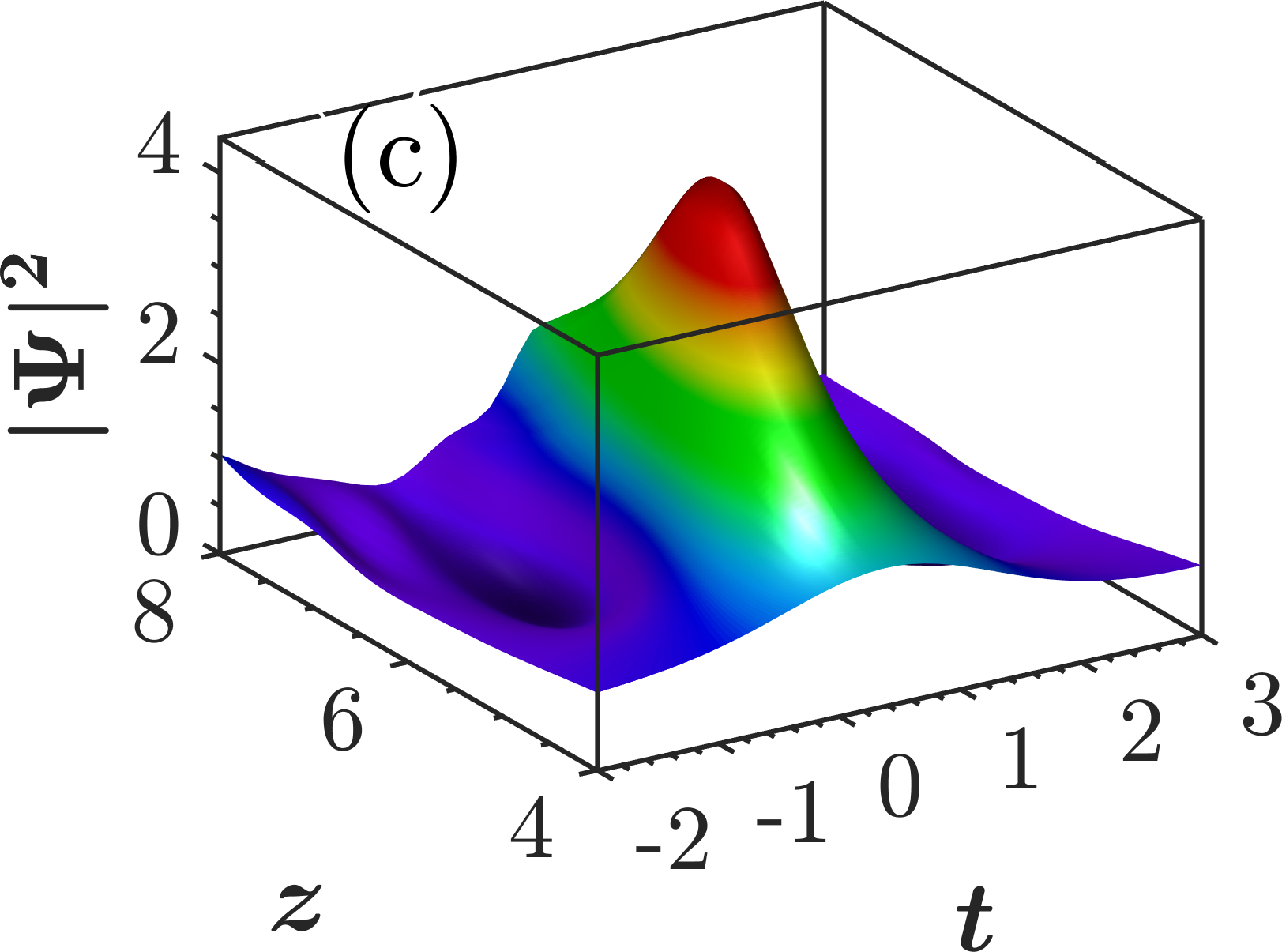}
		\includegraphics[width=0.48\linewidth]{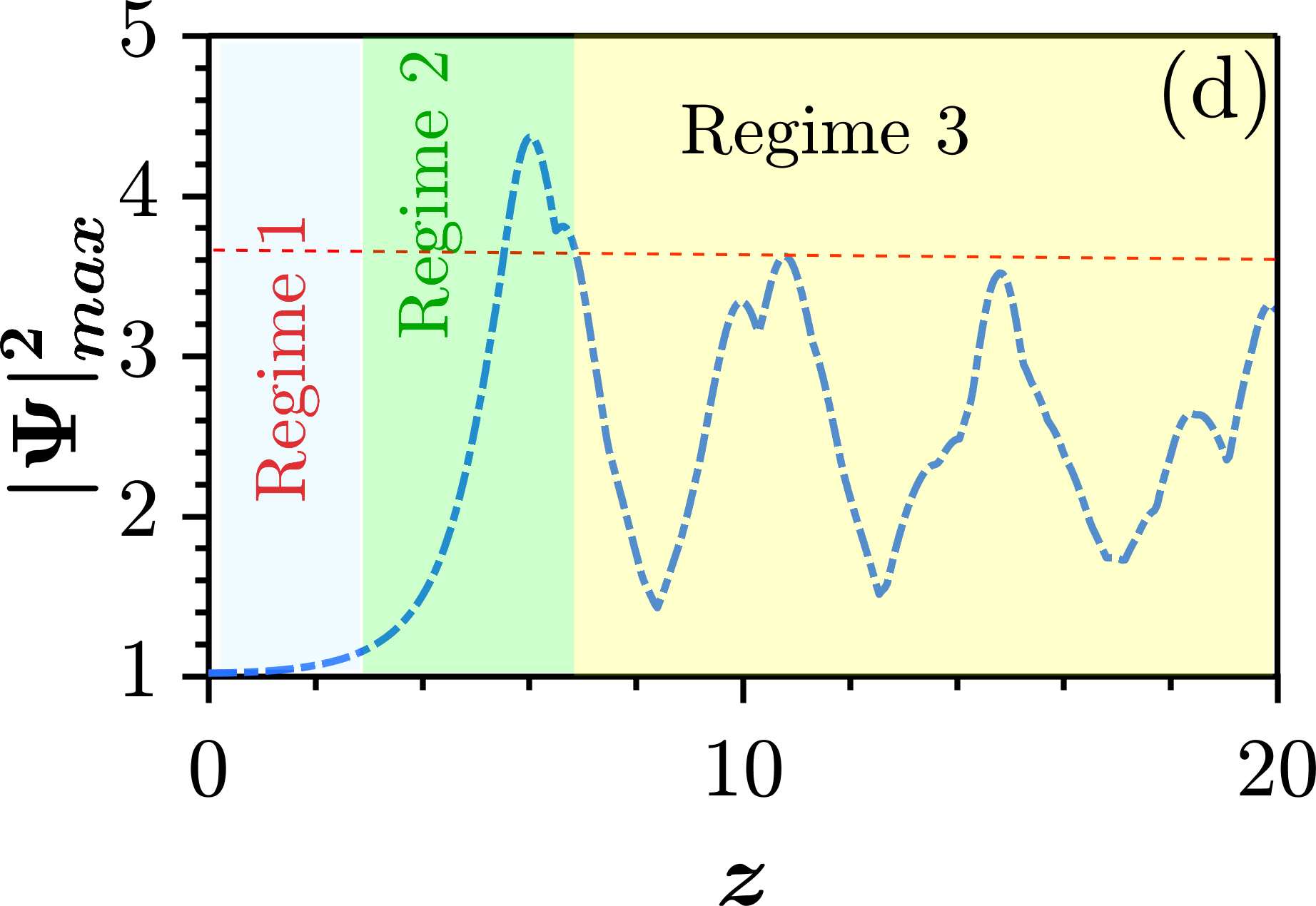}
	\caption{(a) Spatio-temporal evolution of MI of the perturbed plane wave in system with $\Lambda=0.09$ with intensity profiles of (b) AB  and (c) PS (at $z=6$). Growth rate of the MI development along the propagation distance is shown in (d).}
	\label{Numerical}
\end{figure}
\section{Direct numerical evolution of MI dynamics}
To showcase the impact of TOD on the dynamical evolution of ultra-short pulses in the two different physical situations pertaining to \textit{exciton-polariton} with \textit{LHM} and \textit{HEM} cases, we have performed numerical simulations of \eqref{GNLH} by employing the Feit-Fleck algorithm \cite{Feit,New}. For this purpose, we initially choose a plane wave imposed with an infinitesimal perturbation ($\epsilon$) as $\Psi(z=0,t) = A_{0} \left[1 + \epsilon \cos(\hat{\omega} t)\right]$, where $A_0$ refers the amplitude of plane wave, $\hat{\omega}$ is the perturbed wave frequency and  $\epsilon$ takes the value of $10^{-2}$ with $A_0=\hat{\omega}=1$, and $\delta_{3}=0.06$ unless specified.
\subsection{Exciton-polariton with low effective mass}
 We first consider the system with LEM ($\Lambda=0.09$), and the nonlinear development of MI along the propagation distance is shown in Fig.~\ref{Numerical}(a)  corresponding to which (analytical) MI exists. The regular space-time evolution clearly shows the generation of a train of localized pulses with the same magnitude, oscillating periodically in time and localized along $z\sim 5$. These are the typical signatures of Akhmediev breathers (ABs).  The evolution of AB pattern in the vicinity of a sea of pulses is sketched in Fig.~\ref{Numerical}(a). As one progresses along $z$ direction, one can infer intricate splitting of MI peaks as a consequence of energy exchange among sidebands. One can envisage such splitting due to the MI can trigger the formation of localized pulses. Here we notice that the Peregrine soliton (PS) appears in the density MI map as depicted in Fig.~\ref{Numerical}(a). The evolution of PS extracted from Fig.~\ref{Numerical}(a) is shown in Fig.~\ref{Numerical}(c). It is quite remarkable to observe AB and PS by mere shaping of the noise-generated structures through MI stemming from  the TOD in the presence of spatial dispersion in our system \eqref{GNLH}. These results also attest that the AB and PS can provide further insight into the nonlinear localized structures emerging from noise-seeded MI due to the impact of TOD [see Figs. \ref{Numerical}(b) and \ref{Numerical}(c)]. 
To the best of our knowledge, this is the first ever report of higher-order MI for the system with term SD and the possibility of appearance of AB and PS in the underlying system. The corresponding growth rate is shown in Fig.~\ref{Numerical}(d), where it has three specific regions. The  first region lies between $0<z<4$, where the optical wave continues to be a plane wave. Then in the second region, one has symmetric spectral broadening of MI which is nothing but the AB pattern, around $4<z<8$, and the role of TOD leads to a homogeneous shift in the AB pattern. Finally, in the third region, one can observe  the asymmetric nature of the evolution pertaining to the train of ultra-short (soliton) pulses owing to the significant impact of both the TOD and coefficient of SD parameters.
\begin{figure}[t]
	\centering
	\includegraphics[width=0.49\linewidth]{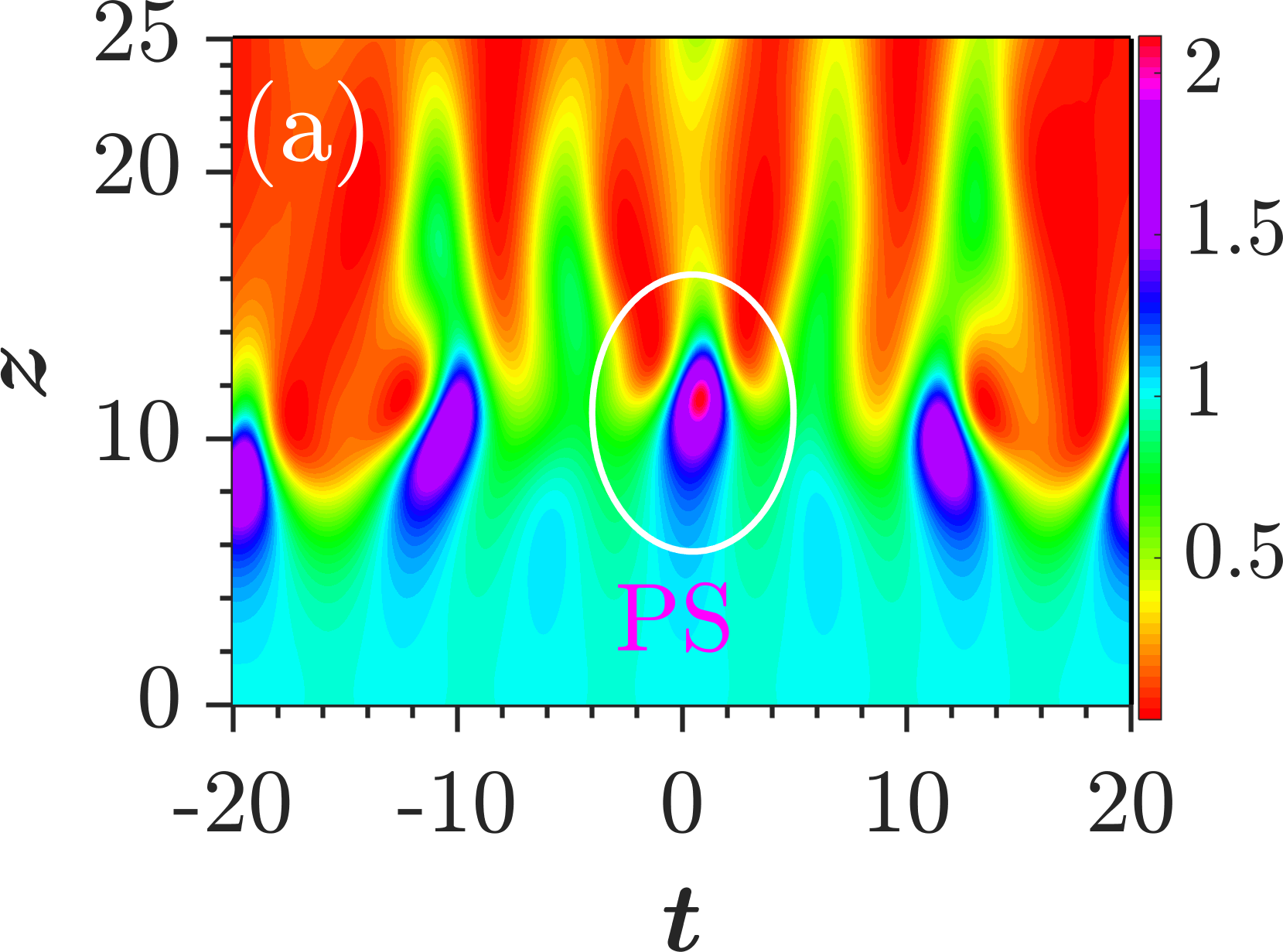}
	\includegraphics[width=0.48\linewidth]{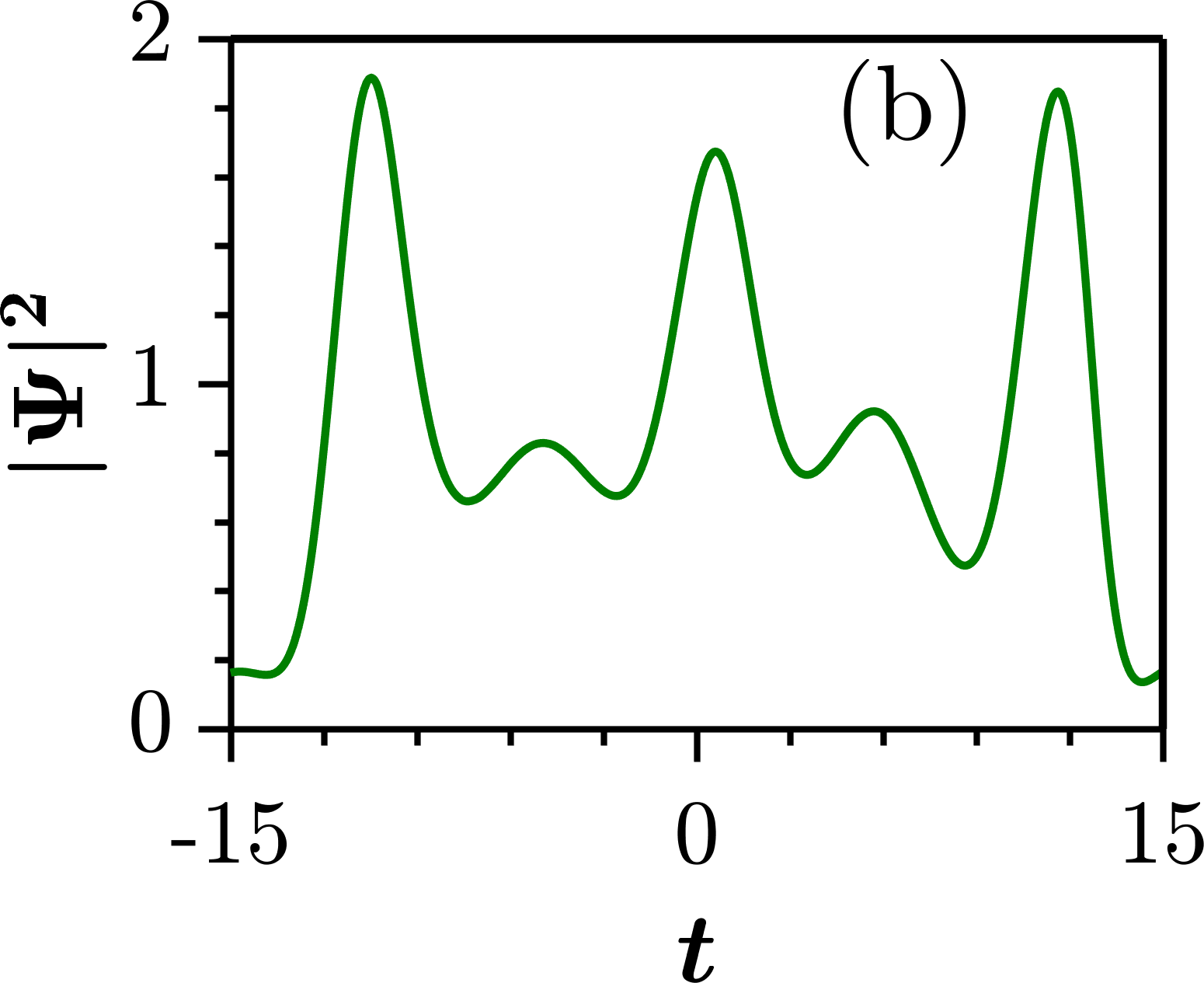}\\
	\includegraphics[width=0.49\linewidth]{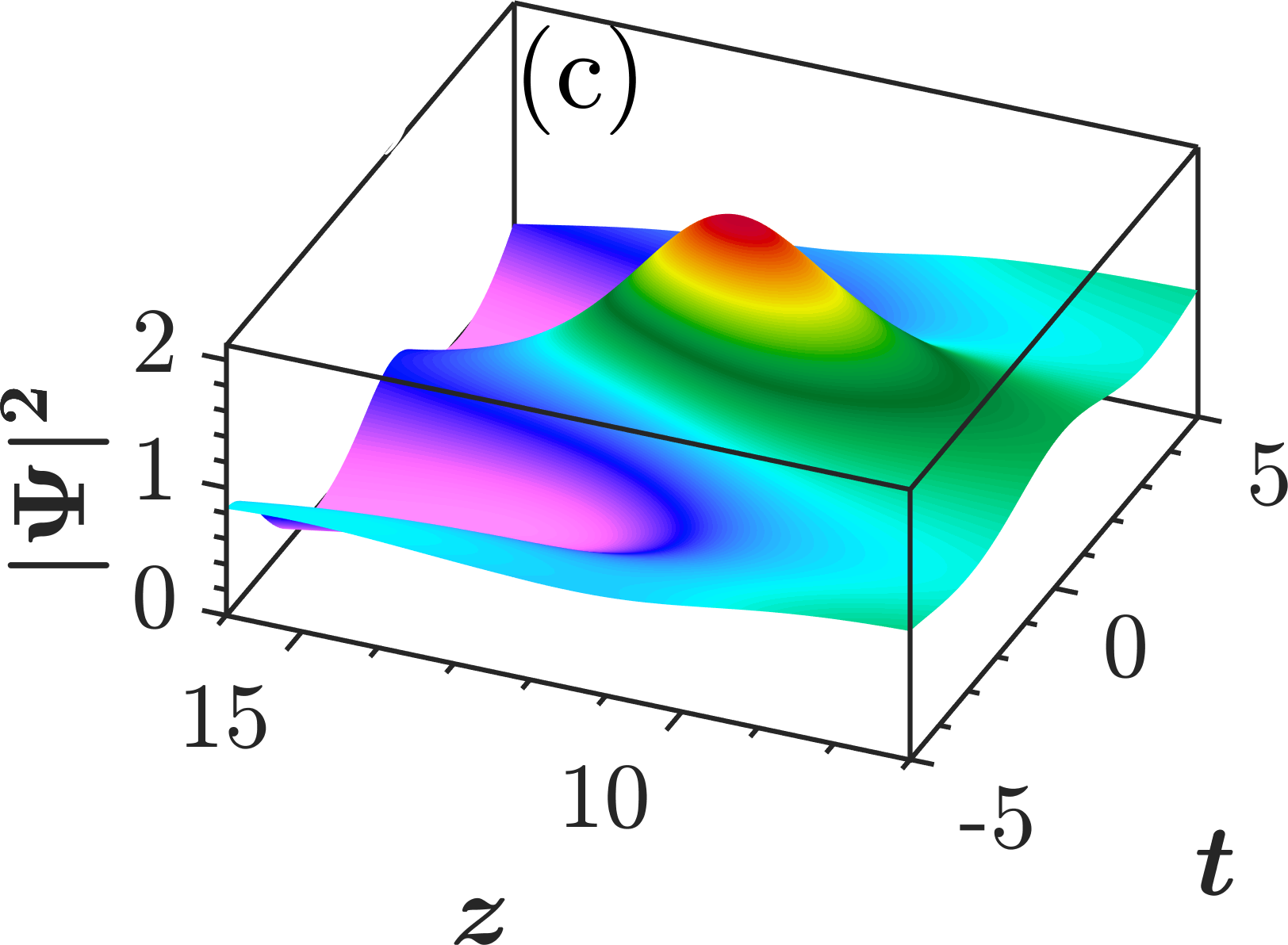}
	\includegraphics[width=0.49\linewidth]{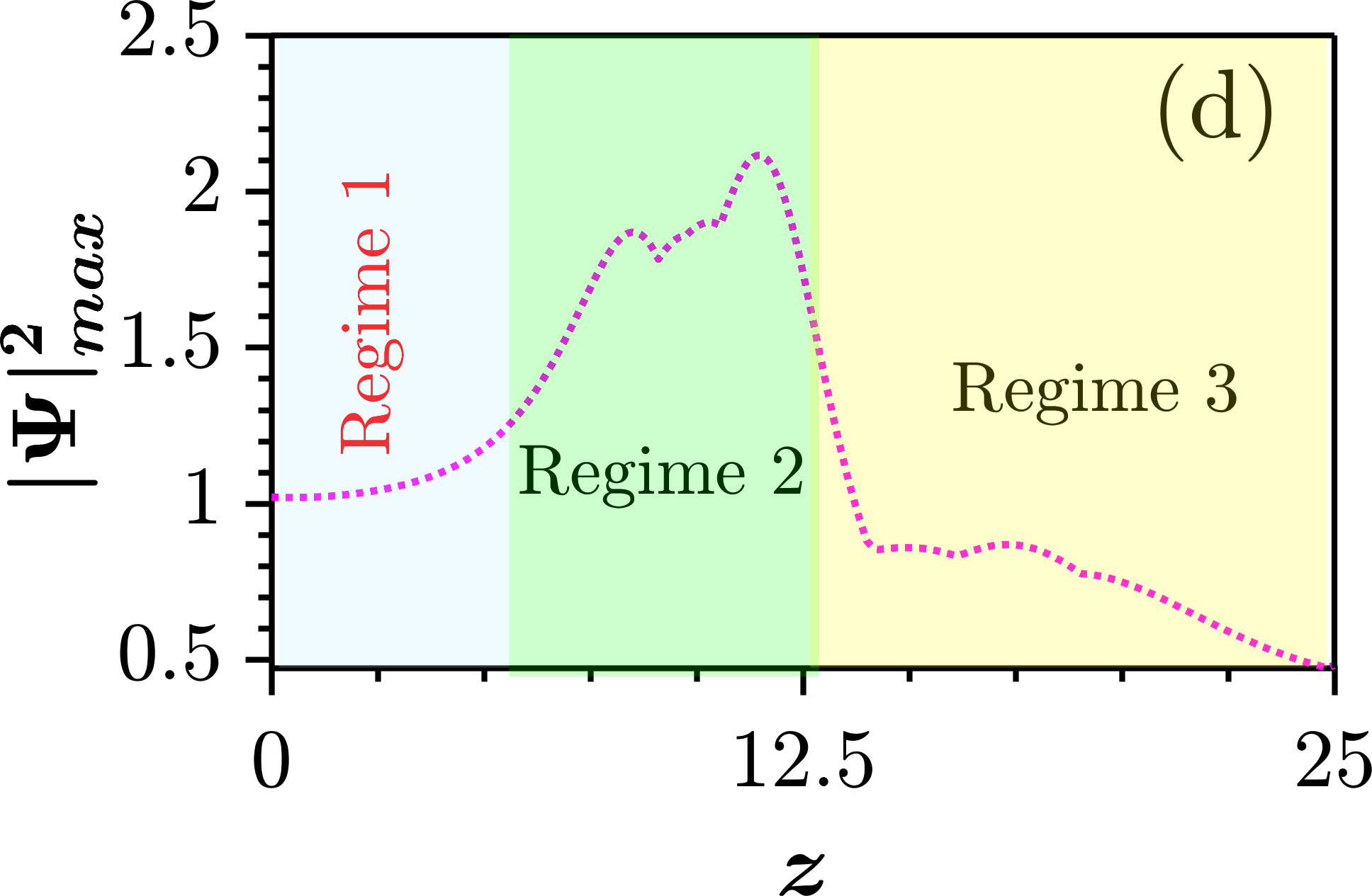}
	\caption{(a) Nonlinear development of MI of the perturbed plane wave in NLS system with $\Lambda=1$. (b) and (c) Portray the two- and three- dimensional views of AB ($z=7$) and PS ($z=10$) extracted from the MI region  marked in Fig. 4(a), respectively. The corresponding growth rate along the evolution coordinate is portrayed in (d).}\label{Numerical1}
\end{figure}
\subsection{ Exciton-polariton with high effective mass}
We finally perform the evolution of complex MI in the physical setting of HEM, where the value of $\Lambda$ can be considered as unity as a consequence of the positive effective exciton mass $M_{ x}^{*}$, correlated with the bulk exciton resonance of Wannier. Unlike the former one, in the case of HEM, we obtain some strange characteristics of MI evolution, where the emergence of first spectral broadening only exists in Fig.~\ref{Numerical1}(a). The first spectral broadening qualitatively resembles a quasi-AB pattern and one can find the PS pattern also in the  central position of AB pattern near $z\sim10$. This is further highlighted in Figs.~ \ref{Numerical1}(b) and \ref{Numerical1}(c), respectively. In Fig. \ref{Numerical1}(d), we display the growth rate of the \textit{exciton-polariton case} which reaches a maximum and then falls down along the $z$-direction indicating no modulation occurs beyond this evolution coordinate, which is then confirmed in Fig. \ref{Numerical1}(a) by manifesting the reduction in the magnitude of  non-zero background.
\section{Conclusion}
To conclude, we have shown that unlike in conventional systems, the TOD can induce a new sort of MI in superlattice system with spatial dispersion indicating the role of Wannier exciton effective mass.  We have also observed that the gain spectrum produced by the TOD is much wider and significantly amplified one compared to the NLS-type spectra. Nonlinear development of MI for the perturbed plane wave has also revealed  the existence of novel ABs and PSs in the system.  
Without being too speculative, we anticipate that these novel ramifications will find applications in both atomic, optical and hydrodynamic extreme events in  higher-order dispersive nonlinear media.

\section*{Acknowledgement}

The work of K T is supported by a Senior Research Fellowship from Rajiv Gandhi National Fellowship (Grant No. F1-17.1/2016-17/R GNF-2015-17-SC -TAM-8989), University Grants Commission (UGC), Government of India. The work of T K was supported by the Science and Engineering Research Board, Department of
Science and Technology (DST-SERB), Government of India, in the form of a major research
project (File No. EMR/2015/ 001408). A G acknowledges the support of DST-SERB for providing a Distinguished Fellowship (Grant No. SB/DF/04/2017) to M L in which A G was a Visiting Scientist. A G is now supported by University Grants Commission (UGC), Government of India, through a Dr. D. S. Kothari Postdoctoral Fellowship (Grant No. F.4-2/2006 (BSR)/PH/19-20/0025).

\end{document}